\documentclass[11pt]{article}

\usepackage{natbib}
\usepackage{graphicx}
\usepackage{txfonts}
\usepackage[normalem]{ulem}
\usepackage{color}
\bibpunct{(}{)}{;}{a}{}{,}

\def\bol{{\rm bol}}
\def\com{{\rm com}}

\def\iso{{\rm iso}}
\def\jet{{\rm jet}}

\def\rr{{\rm r}}
\def\sfr{{\rm SFR}}

\begin{document}

\begin{center}
\noindent
{{\bf\Large Are Gamma-Ray Bursts a Standard Energy Reservoir?}}\end{center}

\begin{center}
 {Li-Xin Li
 }

\vspace{0.3cm}

{\footnotesize{
  Max-Planck-Institut f\"ur Astrophysik, 85741 Garching, Germany
  {\\~~\scriptsize e-mail: lxl@mpa-garching.mpg.de}
  }}
\end{center}

\noindent{\footnotesize {\bf{Abstract.}}
One of the most important discoveries in the observation of gamma-ray 
bursts (GRBs) is that the total energy emitted by a GRB in $\gamma$-rays 
has a very narrow distribution around $10^{51}$~erg, which has led people
to claim that GRBs are standard energy explosions. As people made the 
claim they have ignored the selection biases which must be important since 
GRB observations are strongly fluence or flux-limited. In this paper we 
show that, when the selection effects are considered, the intrinsic 
distribution of the GRB energy can be very broad. The number of faint 
GRBs has been significantly underestimated because of the fluence or flux 
limit. The bright part of the distribution has been affected by another 
important selection effect arising from the beaming of GRB jets, which is 
instrument-independent and caused by the fact that brighter GRBs tend to 
have smaller jet angles and hence smaller probabilities to be detected.
Our finding indicates that GRBs are not a standard energy reservoir, and 
challenges the proposal that GRBs can be used as standard candles to probe 
cosmology.} \\  
\\   
\footnotesize{{\bf Key words:} cosmology: theory -- gamma-rays: bursts -- gamma-rays: observations}

\section{Introduction}
\label{intro}

A characteristic observed feature of cosmological gamma-ray bursts (GRBs) 
is that they emitted a huge amount of energy in $\gamma$-rays in a very 
short time and their isotropic-equivalent $\gamma$-ray energy (i.e., the 
total $\gamma$-ray energy emitted by a GRB if the GRB radiates isotropically) 
spans a very large range---more than 
five orders of magnitude. The measured isotropic-equivalent energy of GRBs, 
$E_\iso$, appears to have a log-normal distribution with a mean $\sim 10^{53}$ 
erg, and a dispersion $\sim 0.9$ in $\log E_\iso$ \citep{ama06,ama07,li07}. 

However, there is evidence that GRBs are beamed \citep{har99,kul99,sta99}.
Assuming that a GRB radiates its energy into two oppositely directed jets,
each having a half-opening angle $\theta_\jet$. The total solid angle
spanned by the jets is then $4\pi\omega$, where $\omega\equiv 1-\cos
\theta_\jet <1$. If the emission of a jet is distributed more or less 
uniformly on its cross-section, the total $\gamma$-ray energy emitted by 
the GRB is approximately $E_\gamma = \omega E_\iso$, smaller than $E_\iso$
by a beaming factor $\omega$. 

One of the most important discoveries in GRB observations has been that the 
value of $E_\gamma$ has a very narrow distribution with a mean $\sim 10^{51}$ 
erg comparable to ordinary supernovae, which has led people to claim that 
GRBs are a standard energy reservoir involving an approximately constant
explosion energy (Frail et al. 2001; Piran et al. 2001; Berger, Kulkarni \& 
Frail 2003; Bloom, Frail \& Kulkarni 2003; Friedman \& Bloom 2005). 
Theoretical models for interpreting the clustering GRB energy have also
been proposed \citep[see, e.g.,][]{zha02}. 

It is well-known that observations of GRBs are strongly fluence or 
flux-limited, hence GRB samples seriously suffer from Malmquist-type selection 
biases \citep{mal20,tee97}. That is, an observer will see an increase in 
the averaged luminosity or the total energy of GRBs with the distance, caused 
by the fact that less luminous or sub-energetic bursts at large distances 
will not be 
detected. Although this Malmquist bias for a flux-limited sample of 
astronomical objects looks obvious, sometimes people made serious mistakes 
in interpreting data by neglecting it. For instance, with a study of nearby 
galaxies it had been incorrectly claimed that the Hubble constant increases 
with the distance \citep{dev72,tee75,san94}.

Unfortunately, as people drew the conclusion on the distribution of the GRB 
energy and claimed that GRBs are standard energy explosions, they have 
treated the observed distribution as the intrinsic distribution and have
neglected the selection biases that are very important for GRBs at 
cosmological distances. As a result, the number of faint GRBs has been 
significantly underestimated, since a GRB will not be detected if its flux 
or fluence falls below the detection limit. The bright part of the GRB 
energy distribution suffers from another important selection bias, which 
arises 
from the fact that brighter GRBs tend to have smaller jet opening angles and 
hence smaller probabilities to be detected. This beaming bias is independent 
of instruments and thus cannot be reduced by improving the sensitivity of
detectors.

The aim of this paper is to show that the influence of the selection biases
from the fluence limit and the jet beaming is strong enough that the observed
distribution of the GRB energy does not represent the intrinsic distribution
at all. We present a simple model that explains nicely the observed 
distribution of the GRB energy, yet the burst energy reservoir in the model
is not standard. Hence, the collimation-corrected energy of GRBs can have a 
very broad intrinsic distribution despite the fact that it is observed to 
cluster to a narrow distribution. Our results lead to the suggestion that 
GRBs are not a standard energy reservoir, contrary to the previous claim.

\section{Intrinsic versus Observed Energy Functions of GRBs}
\label{function}

To estimate the influence of selection biases on the observed distribution
of the GRB energy, we assume that a GRB will be detected if one of its jets
points toward the observer, and its fluence exceeds the limit $F_{\bol,\lim} 
= 1.2\times 10^{-6}$ erg cm$^{-2}$. Although this is an over-simplified 
approximation for the selection effect for GRBs, we will see that this 
simple selection effect already affects the observed distribution of the GRB 
energy strongly enough. In addition, it appears that the above fluence limit
can reasonably represent the selection effect for GRBs with measured peak 
spectral energy and isotropic-equivalent energy \citep{li07}.

The limit in fluence corresponds to a lower limit in the isotropic-equivalent
energy of a detectable GRB at redshift $z$: $E_{\iso,\lim} = 4\pi D_\com^2
(1+z)F_{\bol,\lim}$, where $D_\com$ is the comoving distance to the burst. 
Here we assume a cosmology with $\Omega_{\rm m} = 0.3$, $\Omega_\Lambda=0.7$,
and a Hubble constant $H_0 = 70$ km s$^{-1}$ Mpc$^{-1}$.

For the GRB rate, as people often do we adopt the simplest assumption that 
GRBs follow the cosmic star formation history \citep{tot97,nat05}. Then, up 
to a normalization factor, the intrinsic distribution of GRB redshifts is 
given by
\begin{eqnarray}
	f(z) =  \frac{\Sigma_\sfr(z)}{1+z}\frac{dV_\com}{d z} \;,
	\label{f_sfr}
\end{eqnarray}
where $\Sigma_\sfr(z)$ is the comoving star formation rate, and $V_\com$ is 
the comoving volume.

We adopt a star formation rate \citep{hop06,le07}
\begin{eqnarray}
	\Sigma_\sfr(z) = \frac{1+ a z}{1 + (z/b)^c} \;.  \label{sigma_sfr}
\end{eqnarray}
The parameters $a$, $b$ and $c$ are not well constrained. However, a model 
with $a=8$, $b=3$ and $c=1.3$ fits the observed distribution of GRB 
redshifts reasonably well \citep{le07}. Hence, we fix $a$, $b$ and $c$ to
these values.

For simplicity, we assume that except the number density, the property of 
GRBs does not evolve with the cosmic redshift, although this might not be 
true in reality \citep{li07}. Then, the intrinsic distribution function of 
$z$, $E_\iso$ and $y\equiv\log\left(\tan\theta_\jet\right)$ must have a 
form 
\begin{eqnarray}
	P\left(z,E_\iso,y\right) = f(z) \phi_\iso \left(E_\iso\right) 
		\psi\left(E_\iso, y\right) \;,  
	\label{z_eiso_dist}
\end{eqnarray}
where we assume that $\psi\left(E_\iso, y\right)$ is normalized with respect
to $y$: $\int_{-\infty}^{\infty} \psi\left(E_\iso, y\right) dy = 1$. We
choose $E_\iso$ rather than $E_\gamma$ as an independent variable, since
in practice $E_\iso$ is easier to measure than $E_\gamma$ although the
latter might be more fundamental.

For a GRB with a beaming factor $\omega$, the probability for it to be
detected by an observer is $\omega$, without consideration of the fluence
limit. The observed distribution of $E_\iso$ is then
\begin{eqnarray}
	\hat{\phi}_\iso\left(E_\iso\right) = \phi_\iso\left(E_\iso\right)
		\langle\omega\rangle \left(E_\iso\right) 
		\Xi\left(E_\iso\right)\;,
	\label{hat_phi}
\end{eqnarray}
where
\begin{eqnarray}
	\langle\omega\rangle\left(E_\iso\right) \equiv \int_{-\infty}^{\infty}
		\omega \psi\left(E_\iso, y\right)\, dy
	\label{om_mean}
\end{eqnarray}
is the averaged beaming factor, and the function
\begin{eqnarray}
	\Xi\left(E_\iso\right) \equiv \int_{0}^{z_{\lim}} f(z) dz
	\label{xi}
\end{eqnarray}
reflects the selection effect from the fluence limit.

In equation (\ref{xi}), for a given $E_\iso$, the value of $z_{\lim} = 
z_{\lim}\left(E_\iso\right)$ is solved from the equation $E_\iso = E_{\iso,
\lim}$, or just given by the maximum redshift of GRBs if $E_\iso > E_{\iso,
\lim}$ at $z=z_{\max}$ (assuming that the distribution of GRB redshifts is 
cut off at $z=z_{\max}$).

\begin{figure}[t]
\center{\includegraphics[angle=0,scale=0.49]{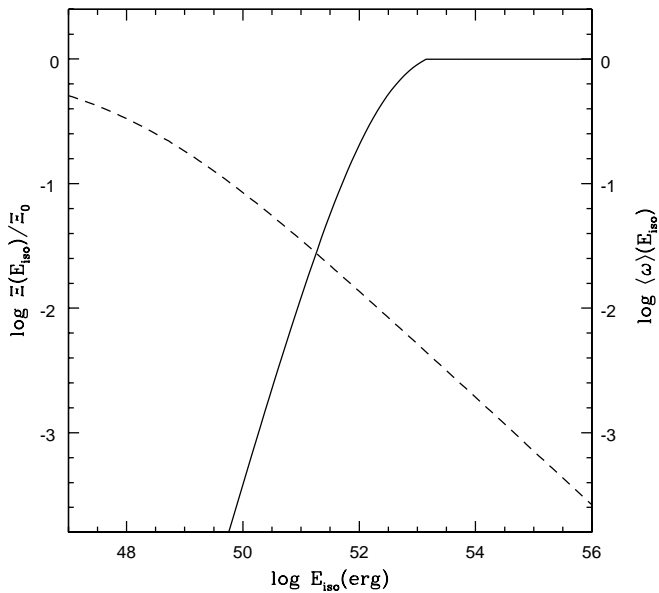}}
\caption{\footnotesize{
The observed distribution of the GRB energy is shaped by the 
fluence limit of the detector and the distribution of the jet opening angle. 
The solid curve shows the fluence-selection function defined by equation 
(\ref{xi}) normalized by $\Xi_0\equiv \int_0^{z_{\max}} f(z) dz$, where we 
have set $z_{\max}=10$. The dashed curve shows the averaged jet beaming 
factor, defined by equation (\ref{om_mean}). The detection criteria are 
defined as follow: a GRB is detected if 
(1) one of its jets points toward the observer; and (2) its observed fluence 
exceeds the limit $F_{\bol,\lim} = 1.2\times 10^{-6}$ erg cm$^{-2}$. 
The fluence-selection function leads to a quasi-exponential cut-off to the 
GRB energy function at the low-energy end. The jet beaming
factor reduces the number of detected GRBs of high energy, caused by the 
fact that brighter GRBs tend to have smaller jet angles.
}}
\label{fig1}
\end{figure}  

\begin{figure}[t]
\center{\includegraphics[angle=0,scale=0.49]{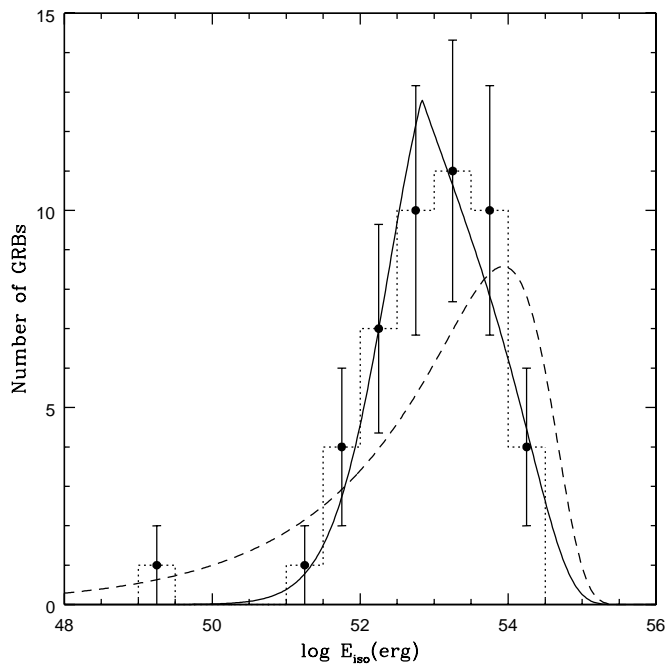}}
\caption{\footnotesize{
The dotted line histogram is the observed distribution of $E_\iso$ 
for 48 long-duration GRBs, with the number of GRBs in each bin indicated by 
a dark point with Poisson error bars. The dashed curve shows the intrinsic 
distribution of $\log E_\iso$ defined by a power law with an exponential 
cut-off in equation (\ref{phi_eiso}), with $\alpha = -0.733$, and $E_\star 
= 3.21\times 10^{54}$ erg. The solid curve is the observed distribution of 
$\log E_\iso$ derived from equation (\ref{hat_phi}), which well fits the 
observed data, with $\chi_\rr^2 = 0.49$. The fluence limit of the detector 
is $1.2\times 10^{-6}$ erg cm$^{-2}$. The maximum GRB redshift is set to
be $z_{\max}= 6$, in accordance with the redshift distribution of the 48 
GRBs. The intrinsic distribution is normalized so that the area under the 
dashed curve is the same as that under the solid curve.
}}
\label{fig2}
\end{figure}

\begin{figure}[t]
\center{\includegraphics[angle=0,scale=0.49]{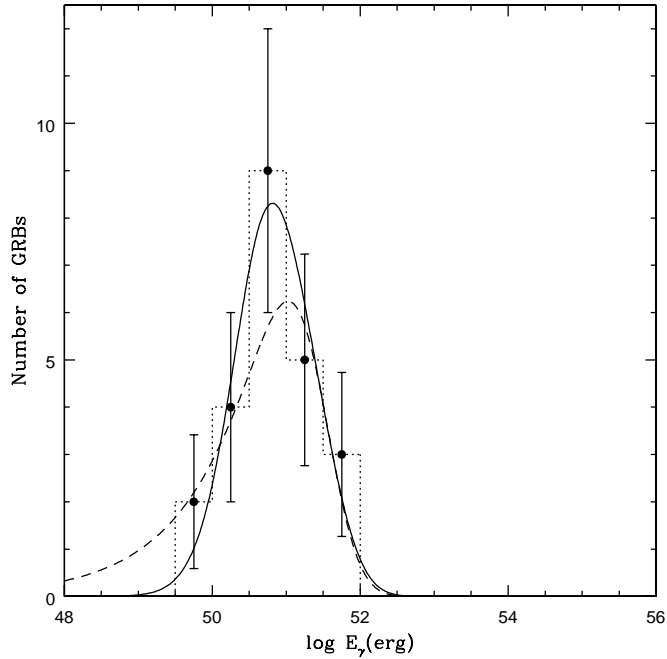}}
\caption{\footnotesize{
Distribution of the collimation-corrected energy of GRBs, 
$E_\gamma$, derived from the intrinsic distribution function of the 
isotropic-equivalent energy in equation (\ref{phi_eiso}), with $\alpha = 
-0.733$ and $E_\star = 3.21\times 10^{54}$ erg (the dashed curve in 
Fig.~\ref{fig2}). The solid curve is the observed distribution of $\log 
E_\gamma$ (eq.~\ref{hat_phig_gamma}). The dashed curve is the intrinsic 
distribution of $\log E_\gamma$ (eq.~\ref{phig_gamma}). The dotted line 
histogram is the distribution of the measured $E_\gamma$ for 23 
long-duration GRBs, with the number of GRBs in each bin indicated by a 
dark point with Poisson error bars. By varying the normalization, the 
derived distribution (the solid curve) fits the observation perfectly, 
with $\chi_\rr^2 = 0.26$. The intrinsic distribution is normalized so that 
the area under the dashed curve is the same as that under the solid curve.
}}
\label{fig3}
\end{figure}

The intrinsic distribution function of the collimation-corrected energy
$E_\gamma = \omega E_\iso$, derived from the distribution in equation 
(\ref{z_eiso_dist}), is
\begin{eqnarray}
	\phi_\gamma\left(E_\gamma\right) =  \int_{-\infty}^{\infty} 
		\omega^{-1}\phi_\iso\left(\omega^{-1}E_\gamma\right) 
		\psi\left(\omega^{-1}E_\gamma,y\right) d y \;.
	\label{phig_gamma}
\end{eqnarray}
The observed distribution of $E_\gamma$ is then
\begin{eqnarray}
	\hat{\phi}_\gamma\left(E_\gamma\right) = \int_{-\infty}^{\infty} 
		\phi_\iso\left(\omega^{-1}E_\gamma\right) \psi\left(
		\omega^{-1}E_\gamma,y\right)\Xi\left(\omega^{-1}
		E_\gamma\right) dy \;.
	\label{hat_phig_gamma}
\end{eqnarray}

It is observed that the jet opening angle of GRBs is anti-correlated to the 
isotropic-equivalent energy \citep{fra01,blo03,fri05}. We define $x\equiv 
\log E_\iso$ and assume that $\psi(x,y)$ has a Gaussian form
\begin{eqnarray}
	\psi(x,y) = \frac{1}{\sqrt{2\pi}\sigma_y} \exp\left[-\frac{(y-
		mx -p)^2}{2\sigma_y^2}\right] \;,
	\label{psi_y}
\end{eqnarray}
where $m$, $p$, and $\sigma_y$ are constants.

For a given $x$, the normalized observed distribution of $y$ is
\begin{eqnarray}
	\left.\hat{\psi}(y)\right|_x = \frac{\omega(y) \psi(x,y)}
		{\int_{-\infty}^{\infty}\omega(y) \psi(x,y)\, dy} 
		= \frac{\omega(y) \psi(x,y)}{\langle\omega\rangle(x)} \;.
	\label{hat_psi}
\end{eqnarray}
Because of normalization, the selection effect from the fluence limit [the
function $\Xi\left(E_\iso\right)$ defined by eq. \ref{xi}] is canceled out in
equation (\ref{hat_psi}) so it does not influence the observed distribution
of the jet opening angle for a given $E_\iso$. However, the selection effect
from beaming [i.e., the function $\omega(y)$] is retained.

A maximum-likelihood fit of equation (\ref{hat_psi}) [with $\psi(x,y)$ given
by eq. \ref{psi_y}] to the 23 GRBs with available $E_\iso$ and $\theta_\jet$ 
\citep{fri05} leads to $m=-0.216$, $p=-0.825$ and $\sigma_y = 0.148$ 
($E_\iso$ in $10^{52}$ erg).

In Fig.~\ref{fig1}, we show the fluence-selection function defined by 
equation (\ref{xi}) (with $z_{\max} = 10$) and the averaged $\omega$ defined 
by equation (\ref{om_mean}). The fluence-selection effect affects the 
observed distribution of faint GRBs, while the beaming effect affects the 
observed distribution of bright GRBs dramatically. For a given intrinsic 
distribution of the GRB energy, the combination of these two effects 
determines the shape of the distribution observed by an observer (if other 
selection effects are neglected). 

Finally, we assume that the intrinsic function of the isotropic-equivalent 
energy of GRBs is a power law with an exponential cut-off
\begin{eqnarray}
	\phi_\iso \left(E_\iso\right) = E_\iso^\alpha \exp\left(-E_\iso
		/E_\star\right) \;,
	\label{phi_eiso}
\end{eqnarray}
where $\alpha$ and $E_\star$ are constant parameters to be determined from
observational data.

\begin{figure}
\center{\includegraphics[angle=0,scale=0.67]{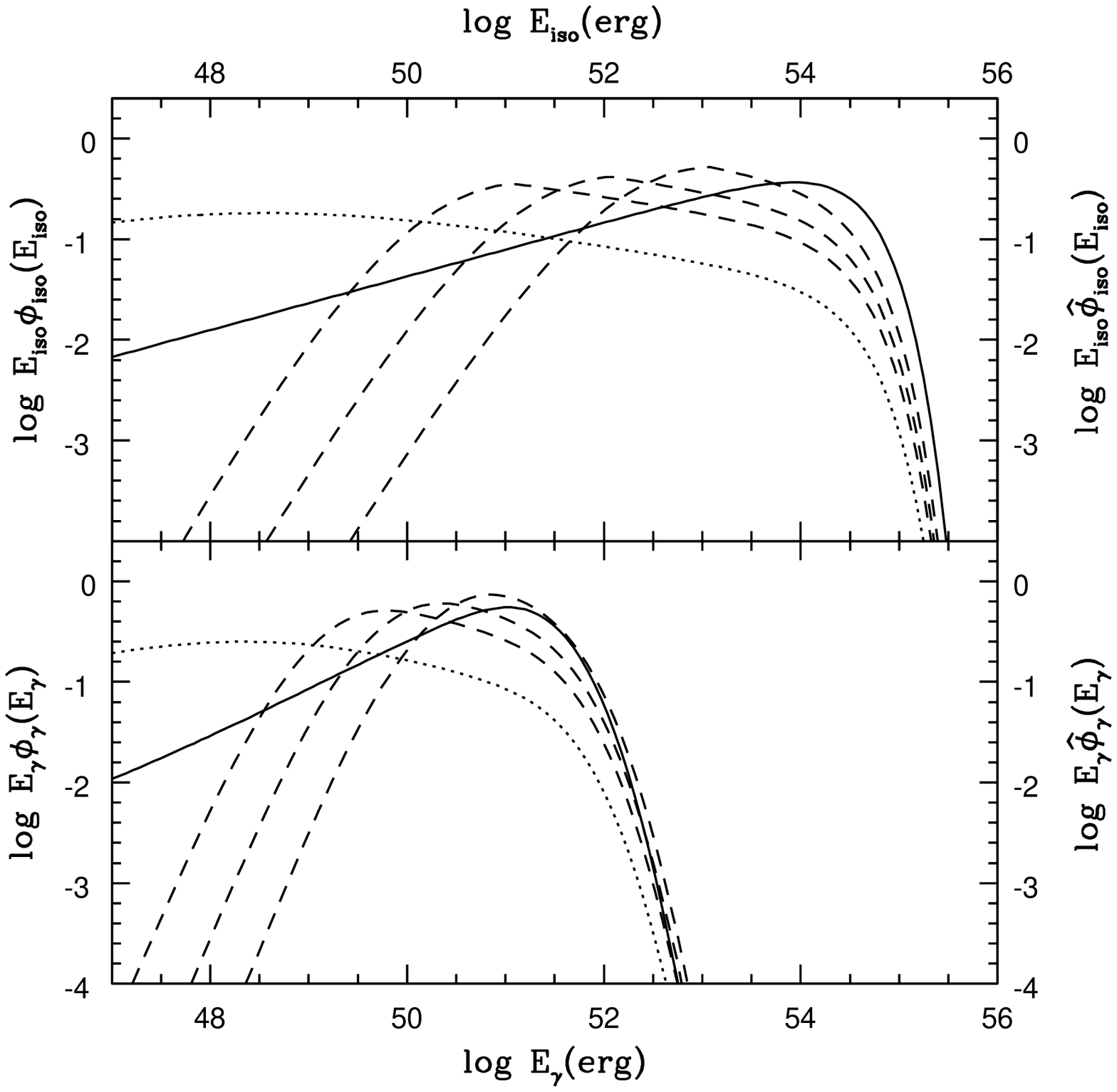}}
\caption{\footnotesize{
This figure shows how the observed distribution of GRB energy
sensitively depends on the selection biases from the detector fluence limit 
and the beaming of GRB jets.
Upper panel:
The intrinsic (solid curve) and the observed (dashed and dotted curves) 
distribution function of the GRB isotropic-equivalent energy, $E_\iso$. The 
solid curve is a plot of $E_\iso\phi_\iso\left(E_\iso\right)$, defined by 
equation (\ref{phi_eiso}) with $\alpha = -0.733$ and $E_\star = 3.21\times 
10^{54}$ erg. The dashed curves are plots of $E_\iso\hat{\phi}_\iso
\left(E_\iso\right)$, calculated by equation (\ref{hat_phi}) with $z_{\max} 
= 10$ and $F_{\bol,\lim} = 10^{-6}$, $10^{-7}$ and $10^{-8}$ erg cm$^{-2}$ 
respectively (from right to left, counted by the peak). The dotted curve 
shows $E_\iso\hat{\phi}_\iso\left(E_\iso\right)$ in the limiting case of 
$F_{\bol,\lim} = 0$ (i.e., GRBs can be detected down to any small value 
of fluence). All distributions are normalized so that the integral over 
$E_\iso$ is unity.
Lower panel:
Similar to the upper panel but for the intrinsic (solid curve) and observed 
(dashed and dotted curves) distribution of the collimation-corrected GRB
energy, $E_\gamma=\omega E_\iso$, calculated by equations (\ref{phig_gamma})
and (\ref{hat_phig_gamma}) with different values of $F_{\bol,\lim}$ as in the
upper panel. All distributions are normalized so that the integral over 
$E_\gamma$ is unity.
}}
\label{fig4}
\end{figure}

\section{Results}
\label{results}

Fitting the observed distribution of $E_\iso$ for 48 long-duration GRBs 
\citep{ama06,ama07,li07} by the function in equation (\ref{hat_phi}), we get 
$\alpha = -0.733$ and $E_\star = 3.21\times 10^{54}$ erg (Fig.~\ref{fig2}). 
The reduced chi-square of the fit is $\chi_\rr^2 = 0.49$, indicating a 
very good fit. This result clearly shows the fact that the intrinsic 
distribution of GRB energy is very different from the observed distribution, 
because of the strong selection effects from beaming and fluence-limit. The
observed distribution has a Gaussian shape, but the intrinsic distribution
is consistent with a power-law with an exponential cut-off.\footnote{We tried 
to fit the observed distribution of $E_\iso$ with a Gaussian intrinsic 
distribution of $\log E_\iso$, but we obtained an unrealistically large mean 
of $E_\iso \sim 10^{60}$ erg.}
The derived intrinsic distribution indicates the existence of a large 
amount of faint GRBs, which have not been detected because of the 
fluence-limit. In addition, the intrinsic distribution $E_\iso\phi_\iso
\left(E_\iso\right)$ peaks at $E_\iso \sim 10^{54}$ erg, an order of 
magnitude larger than the value $\sim 10^{53}$ erg directly inferred from 
the observed distribution.

With $\alpha$ and $E_\star$ fixed at the above values, we then fit the 
observed distribution of $E_\gamma$ for 23 GRBs \citep{fri05} by equation
(\ref{hat_phig_gamma}), varying only the normalization. The best fit is
shown in Fig.~\ref{fig3}, with $\chi_\rr^2 = 0.26$. Although this is not an 
independent fit given the fact that $E_\iso$ and $\theta_\jet$ are 
anti-correlated which has been adopted by our model, the goodness of the fit
in is still impressive, confirming that the relation between $E_\iso$ and 
$\theta_\jet$ assumed in equation (\ref{psi_y}) is a good approximation.

Fig.~\ref{fig3} shows the dramatic difference between the intrinsic
distribution and the observed distribution of $\log E_\gamma$. The observed
distribution (the solid curve) has an exponential decay toward the faint
burst end, but the intrinsic distribution (the dashed curve) decays toward
the faint end by a power law: $E_\gamma \phi_\gamma \left(E_\gamma\right)
\propto E_\gamma^{0.45}$ [slightly faster than the decay of the intrinsic
distribution of $\log E_\iso$, $E_\iso \phi_\iso \left(E_\iso\right)\propto 
E_\iso^{0.27}$]. The existence of a large amount of undetected faint GRBs 
broadens the intrinsic distribution of $\log E_\gamma$ significantly.

Here, by `faint GRBs' we refer to those bursts with an $E_\iso$ or $E_\gamma$
that is smaller than the $E_\iso$ or $E_\gamma$ at the maximum of the
distribution of $\log E_\iso$ and $\log E_\gamma$. Although the detection of
highly sub-luminous and sub-energetic nearby GRBs 980425, 031203 and 060218 
has led people to propose that there exists a unique population of faint 
GRBs \citep{cob06,pia06,sod06,gue07,lia07}, the results in this paper do 
not rely on the existence of this unique population of faint GRBs. The 
extension of the derived energy function of normal GRBs to the low-energy 
end already significantly broadens the GRB energy function.

The effect of the selection biases is more clearly illustrated in 
Fig.~\ref{fig4}, which shows the dependence of the shape of $E_\iso
\hat{\phi}_\iso \left(E_\iso\right)$ (upper panel) and $E_\gamma
\hat{\phi}_\gamma \left(E_\gamma\right)$ (lower panel) on the beaming effect,
and how the shape changes with
the fluence limit of the detector. If the fluence limit decreases by a factor
10 from $10^{-6}$ erg cm$^{-2}$ (or from $10^{-7}$ erg cm$^{-2}$), the width 
of  $E_\iso \hat\phi_\iso \left(E_\iso\right)$ and $E_\gamma \hat\phi_\gamma 
\left(E_\gamma\right)$ increases by $\sim 0.8$ in $\log E_\iso$ and $\sim
0.5$ in $\log E_\gamma$, respectively. Correspondingly, the value of $E_\iso$
at the maximum of $E_\iso \hat\phi_\iso \left(E_\iso\right)$ decreases by a 
factor $\sim 10^{0.8}$, and the value of $E_\gamma$ at the maximum of 
$E_\gamma \hat\phi_\gamma \left(E_\gamma\right)$ decreases by a factor $\sim 
10^{0.5}$.

If we had an ideal detector that can detect an arbitrarily faint GRB,
we would see a distribution of energy given by the dotted lines in 
Fig.~\ref{fig4}, which is dramatically different from the intrinsic 
distribution. For example, the observed distribution of $E_\gamma$ ($E_\iso$)
would peak at $\sim 10^{48}$ erg, rather than $\sim 10^{51}$ erg ($\sim 
10^{54}$ erg) as indicated by the intrinsic distribution.

\section{Conclusions and Discussion}
\label{conclusion}

The observation that the total energy emitted in $\gamma$-rays by 
long-duration GRBs clusters around $10^{51}$ erg \citep{fra01}, 
which has been considered as the most intriguing finding in GRB research 
\citep{zha04}, is only a superficial result since the strong selection biases 
from the detector selection effect and the beaming of GRBs have been ignored.
The previous claim that the energy output of the central 
engine of long-duration GRBs has a universal value \citep{fra01,pir01}, which 
was derived from the above superficial result, is likely to be incorrect 
since the observed narrow distribution of $E_\gamma$ is consistent with a 
broad intrinsic distribution of $E_\gamma$. 

In fact, our results show that for both $\log E_\iso$ (Fig.~\ref{fig2}) and 
$\log E_\gamma$ (Fig.~\ref{fig3}), the distribution on the left-hand side to 
the maximum is well modeled by the cut-off from the fluence limit (the solid 
curve in Fig.~\ref{fig1}). It would be surprising that the intrinsic 
distribution happens to have a low-energy cut-off that is coincident with 
the fluence limit cut-off.

The influence of the flux or fluence limit of detectors on the observation 
of GRBs is well-known and has been taken into account either thoroughly or
partly in many GRB works, e.g. in deriving the luminosity function of GRBs 
(Schmidt 1999, 2001; Firmani et al. 2004; Guetta, Piran \& Waxman 2005; Liang
et al. 2007). However, the influence has sometimes been ignored or seriously 
underestimated. The claim that the 
collimation-corrected energy of GRBs has a narrow distribution and hence
GRBs are a standard energy reservoir is an example where the selection 
effects have been ignored and wrong physical conclusions have been drawn.

Although it is generally conceived that the jet opening angle is 
anti-correlated to the GRB energy, in the study on the luminosity function 
of GRBs the effect of jet beaming has often not been properly taken into 
account. For example, in \citet{gue05} and \citet{lia07}, an isotropic 
luminosity function was derived by comparing the model prediction with the 
observed flux or the luminosity distribution without a
consideration of beaming, then the derived luminosity function was used to 
calculate a weighted and averaged beaming factor. As we can see from 
equation (\ref{hat_phi}), the isotropic luminosity function derived by them
should be the product of the intrinsic isotropic luminosity function and
the averaged beaming factor as a function luminosity, not the intrinsic 
isotropic luminosity itself.

Our results also challenge the proposal that GRBs can be used as standard 
candles to probe cosmology \citep[and references therein]{blo03,fri05,sch07}.
Although the constancy of the GRB energy is not a necessary condition
for GRBs to be standard candles because of the identification of several 
good correlations among GRB observables, the existence of a large amount of
faint bursts that have not been observed might significantly increase the
scatter in those correlations. In addition, the recent work of \citet{but07}
indicates that some of those relations arise from partial correlation with 
the detector threshold and hence are  unrelated to the physical properties 
of GRBs.

Finally, a prediction of this work that can be tested with future 
observations is that as the sensitivity of GRB detectors increases the 
observed distribution of $E_\gamma$ broadens towards the low-energy end.


\end{document}